\documentclass{article} 
\usepackage{iclr2025_conference,times}

\usepackage{amsmath,amsfonts,bm}

\def\eqref#1{equation~\ref{#1}}

\def\1{\bm{1}}

\DeclareMathAlphabet{\mathsfit}{\encodingdefault}{\sfdefault}{m}{sl}
\SetMathAlphabet{\mathsfit}{bold}{\encodingdefault}{\sfdefault}{bx}{n}

\usepackage{hyperref}
\usepackage{url}
\usepackage[inline]{enumitem}
\usepackage{booktabs}
\usepackage{graphicx}
\usepackage{todonotes}
\usepackage{siunitx}
\usepackage{xfrac}
\usepackage{multirow}

\title{A physics-based data-driven model for CO$_2$ gas diffusion electrodes to drive automated laboratories}

\author{Ivan Grega$^*$, F\'elix Therrien$^*$, Abhishek Soni$^\circ$, Karry Ocean$^\circ$, Kevan Dettelbach$^\circ$, \\
\textbf{Ribwar Ahmadi$^\circ$, Mehrdad Mokhtari$^\circ$, Curtis P. Berlinguette$^\circ$, Yoshua Bengio$^{*,\ddagger}$ } \\
$^*$Mila - Quebec AI Institute \hfill $^\circ$University of British Columbia \hfill $^\ddagger$Universit\'e de Montr\'eal \vspace{-3mm}
}

\newcommand*{\CO}{\ensuremath{\mathrm{CO}}}
\newcommand*{\COO}{\ensuremath{\mathrm{CO_2}}}
\newcommand*{\HHO}{\ensuremath{\mathrm{H_2O}}}
\newcommand*{\CCHHHH}{\ensuremath{\mathrm{C_2H_4}}}

\iclrfinalcopy 
\begin{document}

\maketitle

\begin{abstract}
The electrochemical reduction of atmospheric CO$_2$ into high-energy molecules with renewable energy is a promising avenue for energy storage that can take advantage of existing infrastructure especially in areas where sustainable alternatives to fossil fuels do not exist.
Automated laboratories are currently being developed and used to optimize the composition and operating conditions of gas diffusion electrodes (GDEs), the device in which this reaction takes place. Improving the efficiency of GDEs is crucial for this technology to become viable.
Here we present a modeling framework to efficiently explore the high-dimensional parameter space of GDE designs in an active learning context.
At the core of the framework is an uncertainty-aware physics model calibrated with experimental data.
The model has the flexibility to capture various input parameter spaces and any carbon products which can be modeled with Tafel kinetics.
It is interpretable, and a Gaussian process layer can capture deviations of real data from the function space of the physical model itself.
We deploy the model in a simulated active learning setup with real electrochemical data gathered by the AdaCarbon automated laboratory and show that it can be used to efficiently traverse the multi-dimensional parameter space. \vspace{-1mm}
\end{abstract}


    
    


\section{Introduction}


The electrolysis of atmospheric CO$_2$ using renewable electricity is an alternative way to store energy that could allow sectors with limited sustainable alternatives (e.g. aviation, marine, chemical industry) to reach carbon neutrality faster \citep{breyer2019direct, zanatta2023materials}. Electrolyzers take CO$_2$ as input and convert it into valuable chemicals such as carbon monoxide (CO), ethylene (C$_2$H$_4$), and ethanol (C$_2$H$_5$OH) \citep{chen2024catalyst}. This reaction takes place in a part of the device called the gas diffusion electrode (GDE)
whose performance depends on many variables from chemical composition and processing parameters to the choice of the operating conditions, and its optimization is challenging.


Here, we present a physics-based framework for CO$_2$ reduction that can be utilized for the Bayesian optimization of GDE designs in automated self-driving labs. 
The framework uses an analytical one-dimensional continuum model for the GDE cathode which is based on the work by \citet{BLAKE2021138987}. 
We implemented the model in differentiable PyTorch code and extended it to multi-product reactions. \footnote{Code is available at \href{https://github.com/igrega348/CO2-catalysis}{github.com/igrega348/CO2-catalysis}}
Certain latent parameters of the model which cannot be experimentally measured are inferred from the data.
This data-driven analytical model is then used as the mean function of a Gaussian process (GP).
The benefit of this hybrid model is twofold: the GP learns a correction to the physics model, and it provides an uncertainty measure to use in Bayesian optimization.

We demonstrate the use of the model on the experimental data collected from AdaCarbon platform, which consists of a team of seven robots that work together to accelerate the fabrication, characterization and testing of multiple GDEs of varying catalytic compositions. \citep{adacarbon}
As key contributions, we
\begin{enumerate*}[label=(\arabic*)]
    \item  present a physics-based framework for \textit{multi-product} CO$_2$ reduction reactions,
    \item which is interpretable and uncertainty-aware;
    \item demonstrate the use of the model in simulated pool-based active learning with real experimental data.
\end{enumerate*}

\section{Methods}
\paragraph{Data structure and model inputs}
We operate with a GDE dataset collected from a self-driving laboratory, AdaCarbon \citep{adacarbon}. AdaCarbon is a platform comprising seven robotic modules for automated GDE fabrication and characterization, along with an Automated Test Cell capable of performing zero-gap CO$_2$ electrolysis. We used this platform to fabricate and test 90 GDEs (30 unique GDEs, tested in triplicate), with varying compositions of Cu-Ag metals and Nafion-Sustainion ionomer bilayers, aiming to enhance ethylene selectivity.

The available per-sample attributes in the dataset are AgCu ratio \(x_\mathrm{Ag}\), Nafion volume \(v_\mathrm{Naf}\), Sustanion volume \(v_\mathrm{Sus}\) and catalyst mass loading \(m\).
As a pre-processing step, we calculate the thickness of the catalyst layer corresponding to zero porosity as \(L_0=m/(\Bar{\rho}A)\) where \(\Bar{\rho}\) is the mass-averaged density given by
\begin{align*}
    \Bar{\rho} &= (1-x_\mathrm{Ag})\rho_{\mathrm{Cu}} + x_\mathrm{Ag}\rho_{\mathrm{Ag}}
\end{align*}
This way we obtain the set of 5 input variables \(x_\mathrm{Ag},v_\mathrm{Naf},v_\mathrm{Sus},m,L_0\).
There are two output variables: the Faradaic efficiencies with respect to {\CO} and ethylene, \(\mathrm{FE_{CO}}\) and \(\mathrm{FE_{\CCHHHH}}\), respectively.

In addition to the per-sample attributes, there are further parameters which characterize the operating conditions of the CO2 reduction system which is run at steady-state conditions. 
The values of these parameters and other assumed physical constants are listed in the Appendix.
Notably, the predictions of the model are very sensitive to certain parameters which are not readily available from the experiments (e.g. porosity of the cathode \(\varepsilon\), radius of particles \(r\), diffusion coefficient \(K_{dl}\)).
We employ an MLP with layer dimensionality \([5,64,64,64,6]\),  ReLU nonlinearities and dropout with \(p=0.1\) between the layers to map from the five per-sample input variables to the six per-sample inputs to the physical model (\(\varepsilon, r, K_{dl}, \theta_1, \theta_2, \theta_3\)).
In addition to the per-sample inputs, there are 6 trainable parameters in the model which do not depend on sample attributes. 
These are the Tafel reaction constants \(i_i^*,\alpha_i\) which are defined in the Appendix.

\begin{figure}
    \centering
    \includegraphics[width=\linewidth]{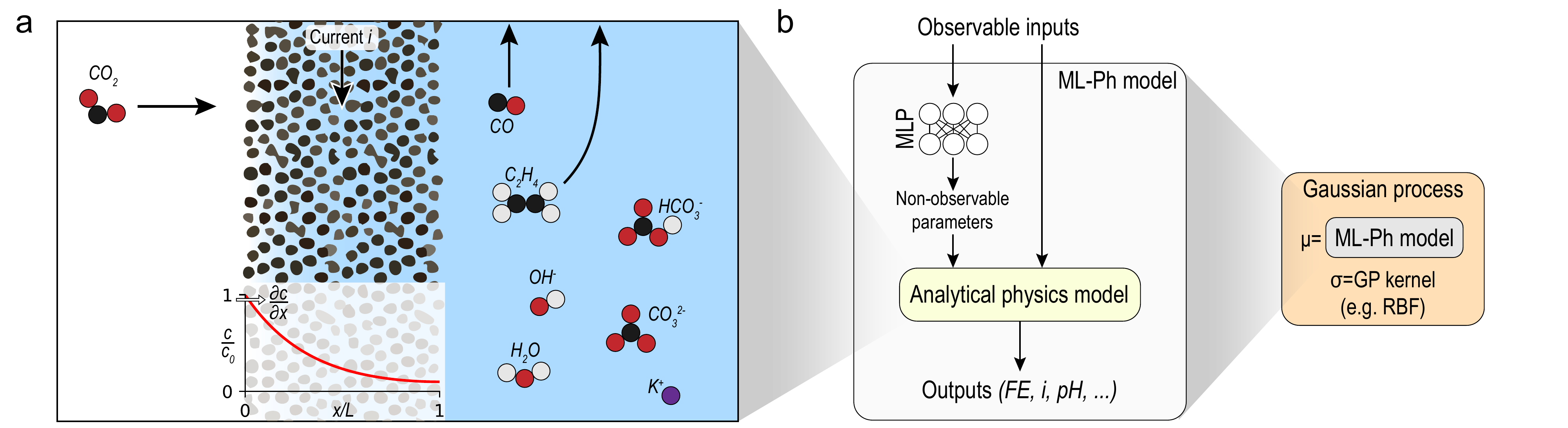}
    \caption{\textit{a)} Schematic of the analytical model of the cathode. {\COO} dissolves in the electrolyte as it enters the porous catalyst layer. The concentration profile of {\COO} \((c/c_0)\) is indicated on the schematic. Gaseous reaction products leave the system while other molecules and ions are dissolved in the electrolyte and in equilibrium.
    \textit{b)} The analytical model is embedded in a data-driven framework (ML-Ph) where non-observable parameters are inferred from the data. 
    \textit{c)} The ML-Ph model is used as the mean function of a Gaussian process to enable uncertainty-aware predictions.
    }
    \label{fig:Fig1}
\end{figure}

\paragraph{Differentiable inverse solver}
In the basic operation, the model takes the cathode voltage along with the inputs as mentioned above and solves for concentrations of products, current densities and the associated Faradaic efficiencies.
In our experiments, however, it is the cathode current rather than cathode voltage which is imposed.
Moreover, in the current experimental setup the cathode voltage is not measured (only the voltage across the entire cell could be measured).
Therefore, 
given the current state of the model parameters and inputs, the outputs are calculated for a set of 1000 voltages ranging between \SI{-1.25}{V} and \SI{0}{V}.
We then find the operating point as a linear interpolation of the 2 voltages which provide the overall current closest to the target value of \(I=\SI{233}{mAcm^{-2}}\).
This procedure is efficient, enables backpropagation of gradients to the MLP, and proves empirically stable.

\paragraph{Uncertainty-aware model}\label{sec:uncertainty-aware-model}
In order to use the model in Bayesian optimization of experiments, every prediction needs to come with an associated uncertainty measure.
We experiment with two frameworks: 
\begin{enumerate*}[label=(\roman*)]
    \item model ensembling and
    \item Gaussian process (GP) with embedded physics mean function.
\end{enumerate*}
The ensembles are trained with a bagging approach whereby each of the \(N=50\) models has access to a fraction \(m=\sfrac{1}{3}\) of the data during training.
At inference time, predictions are obtained from each model independently and variance of the predictions is used as the resulting uncertainty.
In the GP with embedded physics mean function, a single physics model is used as the mean function and GP covariance kernel is trained on top of it to provide an uncertainty measure at inference time.

\section{Results}

\subsection{Uncertainty-aware predictions}
As outlined in Section~\ref{sec:uncertainty-aware-model}, we experiment with two approaches:
\begin{enumerate*}[label=(\roman*)]
    \item model ensembling (Ph ensemble) and
    \item Gaussian process with physics mean function (GP + Ph).
\end{enumerate*}
In addition, we consider two other models:
\begin{enumerate*}[label=(\roman*)]
    \item ensemble of MLPs (MLP ensemble) and
    \item constant-mean Gaussian process (GP).
\end{enumerate*}
Both these models are trained end-to-end with 5 inputs (\(x_\mathrm{Ag},v_\mathrm{Naf},v_\mathrm{Sus},m,L_0\)) and 2 outputs (Faradaic efficiency with respect to carbon monoxide and ethylene, \(\mathrm{FE}_{\CO}\) and \(\mathrm{FE}_{\CCHHHH}\), respectively).

\begin{figure}
    \centering
    \includegraphics[width=\linewidth]{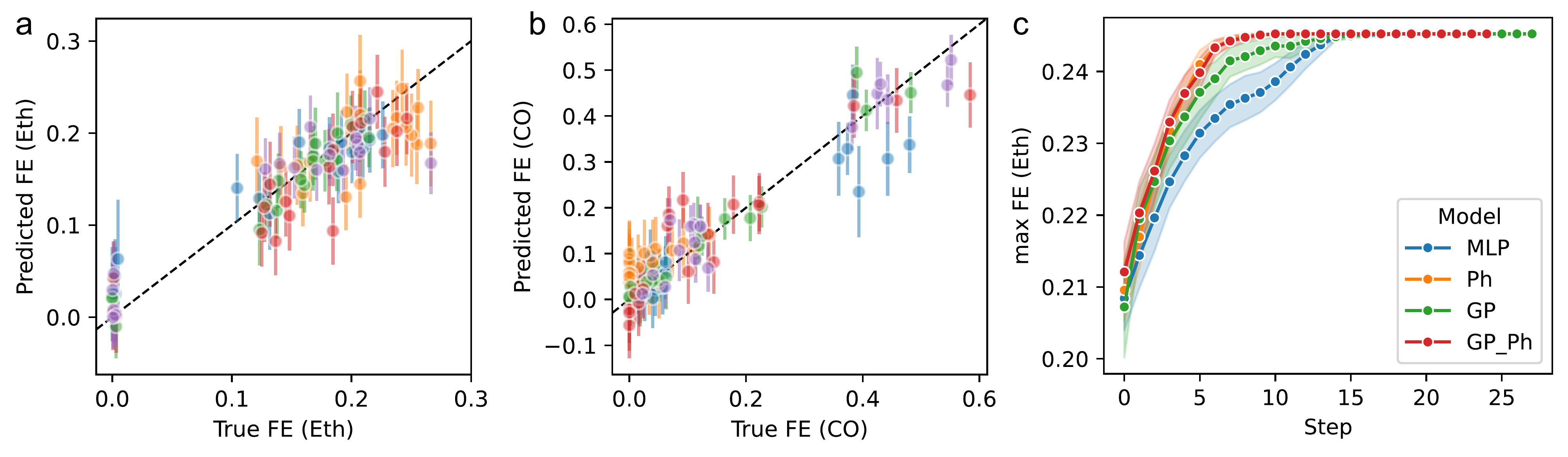}
    \caption{
    \textit{a,b)} Parity plots over test set for Faradaic efficiency of ethylene and CO, respectively for the GP+Ph model. Error bars are \(\pm \sigma\). Colors represent independent trials in 5-fold cross-validation.
    \textit{c)} Maximum FE in the training pool as a function of step number in simulated active learning.
    }
    \label{fig:Fig-fit-AL}
\end{figure}

Table~\ref{tab:fit-4-models} shows the negative log likelihood and mean average error for the two-target predictions of both {\COO} and {\CCHHHH} Faradaic efficiencies. 
Of the four evaluated approaches, the Gaussian process with embedded physics mean function performs best on both metrics.
Figure~\ref{fig:Fig-fit-AL}a,b shows the parity plots over withheld test set for the GP+Ph model. 
In all five runs of the 5-fold validation, the predicted values are close to the true values and the predicted uncertainty is well calibrated.

\begin{figure}[b]
    \centering
    \includegraphics[width=\linewidth]{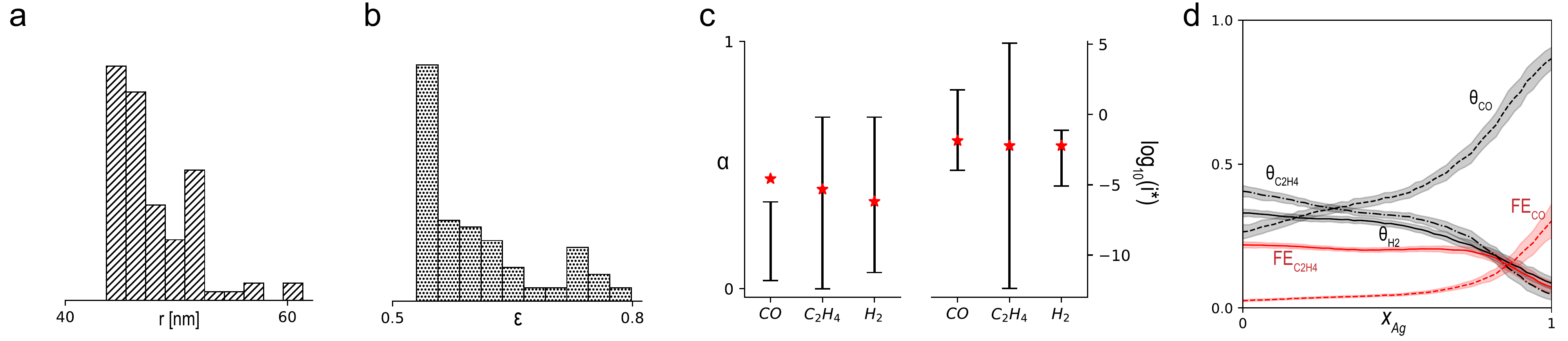}
    \caption{
        \textit{a,b)} Histograms of inferred particle radius \(r\) and porosity \(\varepsilon\) for our dataset.
        \textit{c)} Electrode reaction kinetics parameters \(\alpha\) and \(i^*\). Vertical bars show ranges of values reported in the literature \citep{lees2022} on \textit{Cu} substrates.
        Red stars show values inferred for our dataset (mixture of {Cu} \textit{and} {Ag} substrates).
        \textit{d)} Faradaic efficiencies and surface coverage parameters \(\theta_i\) for a virtual \(x_\mathrm{Ag}\) sweep. Shaded bands indicate dropout-enabled variance \citep{gal2016dropout}.
    }
    \label{fig:Fig-interpretability}
\end{figure}

\subsection{Interpretability of the model}
Certain non-observable parameters of the model (e.g. porosity \(\varepsilon\), particle radius \(r\)) are per-sample while others (e.g. electrode reaction kinetics) are shared across samples. Characteristics of the inferred parameters are displayed in Fig.~\ref{fig:Fig-interpretability}a,b.
Values of particle radius \(r\) and porosity \(\varepsilon\) are in sensible ranges (\(\SI{40}{nm} < r < \SI{65}{nm}\), \(0.5 < \varepsilon < 0.8\)).
In Fig.~\ref{fig:Fig-interpretability}c we plot the inferred values of electrode reaction kinetics \(\alpha_i\) and \(i^*_i\) and we compare them to the values reported in the literature \citep{lees2022}.
Note that unlike reaction potentials \(\eta_i\) which are physical constants depending on the molecule formation energies, there is a large variability in values of \(\alpha_i\) and \(i^*_i\) reported in the literature.
Nonetheless, our inferred values fall within the acceptable ranges.\footnote{The plotted range of reported values of \(\alpha\) corresponds to copper cathodes while our cathodes are a mixture of copper and silver. Reported values of \(\alpha_\CO\) on silver cathodes are higher.}

Finally, the analytical model allows for physical interpretability of internal parameters.
In a ``virtual experiment'', we vary the AgCu ratio while keeping all other variables at their mean values. 
The Faradaic efficiencies change accordingly and we observe that internally the inferred per-sample values of surface coverages \(\theta_i\) indicate that reactions switch from preferential ethylene production on copper-rich surfaces (\(\theta_{\CCHHHH}>\theta_{\CO}\) at \(x_\mathrm{Ag}\sim0\)) to carbon monoxide production on silver-rich surfaces (\(\theta_{\CCHHHH}<\theta_{\CO}\) at \(x_\mathrm{Ag}\sim1\)) (Figure~\ref{fig:Fig-interpretability}d).

\begin{table}
    \centering
    \caption{Negative log likelihood (NLL), mean average error (MAE) and acceleration factor (AF) for the benchmark of 4 models.\vspace{2mm}}
    \resizebox{0.5\textwidth}{!}{%
    \begin{tabular}{ccccc}
        \toprule
        {} & Model & NLL \(\downarrow\) & MAE \(\downarrow\) & AF \(\uparrow\) \\
        \midrule
         \multirow{2}{*}{Benchmark} & MLP ensemble & \(-1.4 \pm 0.3\) & \(0.045 \pm 0.02\) & \(2.0\) \\
         {} & GP & \(-1.5 \pm 0.4 \) & \(0.048 \pm 0.01\) & \(2.9\) \\
         \midrule
         \multirow{2}{*}{Ours} & Ph ensemble & \(-1.5 \pm 0.3\) & \(0.048 \pm 0.01\) & \(\bm{3.2}\) \\
         {} & GP + Ph & \(\bm{-1.9 \pm 0.3}\) & \(\bm{0.025 \pm 0.01}\) & \(3.1\) \\
         \bottomrule
    \end{tabular}
    }
    \label{tab:fit-4-models}
\end{table}

\subsection{Simulated pool-based active learning}
We carry out simulated active learning for maximization of ethylene Faradaic efficiency.
All available data points are split into 2 pools: training pool and candidate pool. 
The models are calibrated using the data from the training pool and the next candidate is proposed from the candidate pool based on the acquisition function (here \emph{expected improvement}).
Once a data point is selected, it is included in the training pool for the next round of model calibration. See Appendix for more details.

In Figure~\ref{fig:Fig-fit-AL}c we plot the maximum Faradaic efficiency over the training pool as a function of acquisition step. 
The shaded bands correspond to variability over 100 independent trials.
It is clear that the models with embedded physics perform better than the pure GP while an MLP ensemble has the worst performance.
In Table~\ref{tab:fit-4-models} we include the \textit{acceleration factor} (AF) with respect to random sampling. 
The ensemble of MLPs is on average \(2\times\) more efficient than random sampling while the other three models are approximately \(3\times\) more efficient.

\section{Current limitations and outlook}
In this work, we developed a physics-based data-driven model for gas diffusion electrodes where non-observable parameters are inferred from experimental data.
We have shown that the model can capture multi-product reactions with CO and ethylene products, but it is easily extensible to other carbon products.
We carried out simulated pool-based active learning to evaluate the capacity of the framework to guide automated experiments. 
The Gaussian process (GP) augmented with a physics model achieves best performance.
While the performance gain over standard GP is small, more experiments are needed to verify how the models perform in real scenarios with continuous optimization domain.
Moreover, the physics-based model provides physical understanding and explainability not found in black-box ML models.
In this work we focused on optimization deployed on a specific robot, but we are working on expanding the capabilities for other systems.


\subsubsection*{Acknowledgments}
This project was supported by an NSERC Alliance Grant (ALLRP 577135-2022).

\bibliography{iclr2025_conference}
\bibliographystyle{iclr2025_conference}

\appendix
\section{Appendix}
\subsection{Sensitivity analysis}
\begin{figure}
    \centering
    \includegraphics[width=0.75\linewidth]{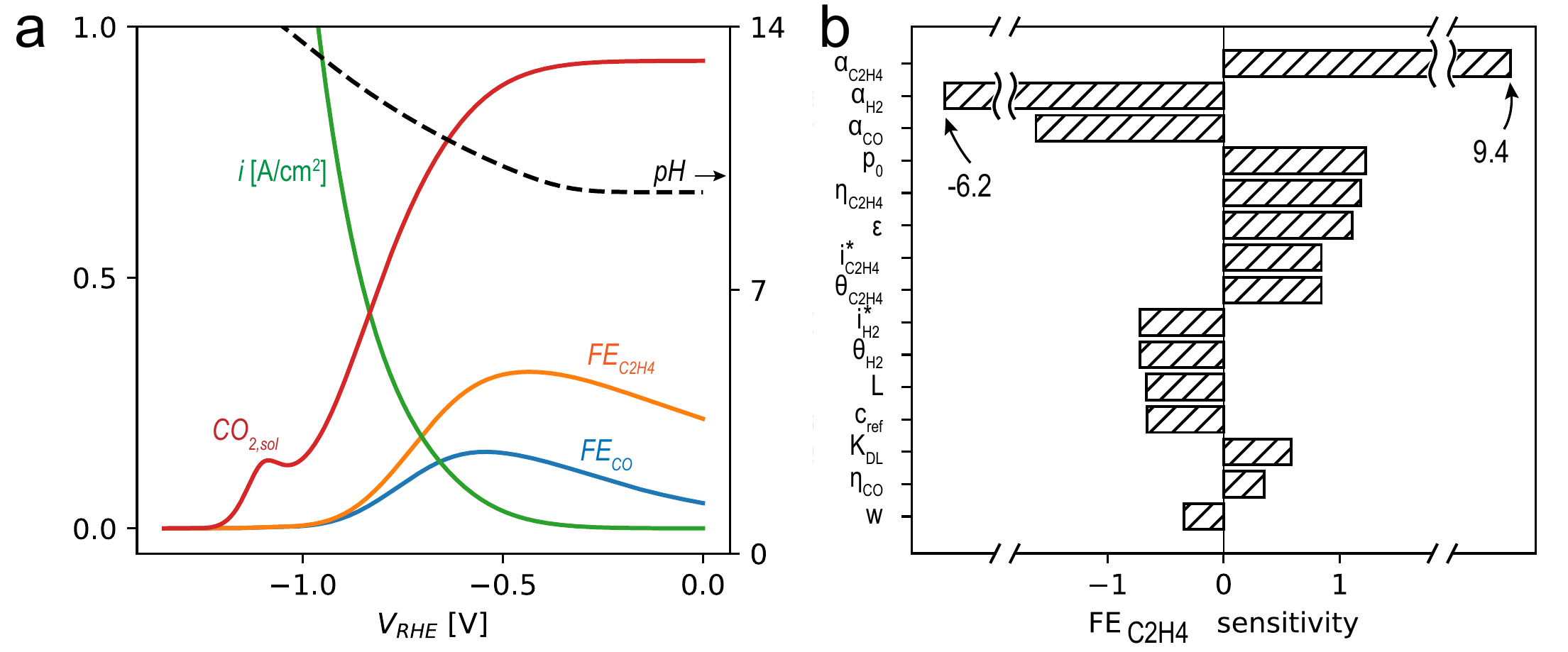}
    \caption{
        \textit{a)} Dependence of selected quantities on half-cell voltage \(V_{\mathrm{RHE}}\).
        \textit{b)} Sensitivity of ethylene Faradaic efficiency to model parameters.
    }
    \label{fig:Fig-app-model}
\end{figure}

The analytical model provides a detailed view of the micro-environment at the cathode. 
In Figure~\ref{fig:Fig-app-model}a we plot the dependence of various parameters on the half-cell voltage with respect to the reversible hydrogen electrode \(V_\mathrm{RHE}\) for a representative set of parameters.
As cathode voltage becomes increasingly more negative, the overall cathode current \(i\) increases exponentially (as governed by Tafel kinetics).
Faradaic efficiencies \(\mathrm{FE}_{\CCHHHH}\) and \(\mathrm{FE}_{\CO}\) first increase but then begin to drop again below \(V_\mathrm{RHE}\sim-0.5\). 
This behavior is well known in the literature \citep{BLAKE2021138987} and is due to the excessive production of OH ions from hydrogen evolution, the environment becoming highly alkaline (see the associated increase in pH) and
the associated reduction in the solubility of \COO in the electrolyte, \(\mathrm{CO}_{2,sol}\).

We now vary the input parameters of the model and observe the associated change in the output Faradaic efficiency for ethylene, \(\mathrm{FE}_\CCHHHH\) (Fig.~\ref{fig:Fig-app-model}b).
Sensitivity of \(\mathrm{FE}\) to parameter \(x\) is defined as 
\begin{equation*}
    \frac{\frac{\partial \mathrm{FE}}{\mathrm{FE}}}{\frac{\partial x}{x}}={\frac{\partial \mathrm{FE}}{\partial x}}{\frac{x}{\mathrm{FE}}}
\end{equation*}
where the derivative \(\frac{\partial \mathrm{FE}}{\partial x}\) is estimated numerically by taking small deviations around \(x\).
The parameters with largest sensitivity are electrode kinetics parameters, \(\alpha_i\).
For instance, a 1\% change in \(\alpha_{C2H4}\) will result in 9.4\% change in \(\mathrm{FE}_\CCHHHH\).
An important observation is that many parameters which are not readily observable in experiments (e.g. \(\alpha_i, i^*_i, \theta_i,\varepsilon,K_{DL}\)) have a deciding impact on the output Faradaic efficiency.
For this reason, we implement a data-driven approach in which we infer these latent parameters from the data.

\subsection{Details of the physics model}
The following three reactions are assumed to take place at the cathode:
\begin{align}
     2\HHO + 2\mathrm{e^-} &\longrightarrow \mathrm{H_2} + 2\mathrm{OH^-} \\
     \COO + \HHO + 2\mathrm{e^-} &\longrightarrow \CO + 2\mathrm{OH^-} \\
     2\COO + 8\HHO + 12\mathrm{e^-} &\longrightarrow \mathrm{C_2 H_4} + 12\mathrm{OH^-}
\end{align}
The corresponding current densities are expressed using Tafel kinetics with first order dependence on the local concentration of reactants and exponential dependence on overpotentials \(\eta_i\):
\begin{align*}
     i_{H_2} &= -\theta_1 i_{H_2}^* \exp{\left( -\frac{\alpha_{H_2} F }{RT} \eta_{H_2} \right)} \\
     i_{CO} &= -\theta_{CO} i_{CO}^* \frac{c_{\COO}}{c_\mathrm{ref}} \exp{\left( -\frac{\alpha_{CO} F }{RT} \eta_{CO} \right)} \\
     i_{C_2H_4} &= -\theta_{C_2H_4} i_{C_2H_4}^* \frac{c_\COO}{c_\mathrm{ref}} \exp{\left( -\frac{\alpha_{C_2H_4} F }{RT} \eta_{C_2H_4} \right)}
\end{align*}
The factors \(i_{H_2}^*,i_{CO}^*,i_{C_2H_4}^*,\alpha_{H_2},\alpha_{CO},\alpha_{C_2H_4}\) are trainable constants. 
As shown in review by \citet{lees2022}, the range of reported constants \(i_i^*\) and \(\alpha_i\) varies by orders of magnitudes between different works in the literature.
Therefore, we argue that it is reasonable to enable these constants to be trained from data.
The competition between the three reactions is captured by \textit{coverage parameters } \(\theta_{H_2}, \theta_{CO}, \theta_{C_2H_4}\).
These are similar to the surface coverage parameters outlined in \citep{lees2022} and introduced by \citet{langmuir1918}. 
However, we introduce a different framework in which we do not calculate \(\theta_i\) from temperature and voltage, but rather 
\(\theta_i\) are functions of the input parameters (e.g. AgCu ratio). 
They sum to 1 which we enforce by softmax.
For instance, as shown in Figure~\ref{fig:Fig-interpretability}, the model trains to shift the balance of the reactions to ethylene on Cu-rich surfaces and to {\CO} on cathodes without copper.

\subsubsection{Physical constants}
The nominal surface area of the cathode is \(A=(\SI{1.85}{cm})^2\), the overall current density is \(i=\SI{233}{mA cm^{-2}}\), {\COO} is fed from a fixed pressure reservoir with \(p_0=\SI{2.38}{bar}\) at a flow rate of \(Q=\SI{30}{cm^3/min}\). Flow channel dimensions are \(\SI{1.5}{mm}\times\SI{0.5}{mm}\times\SI{20}{mm}\).

\subsection{Calibration of model ensembles}
\begin{figure}
    \centering
    \includegraphics[width=0.9\linewidth]{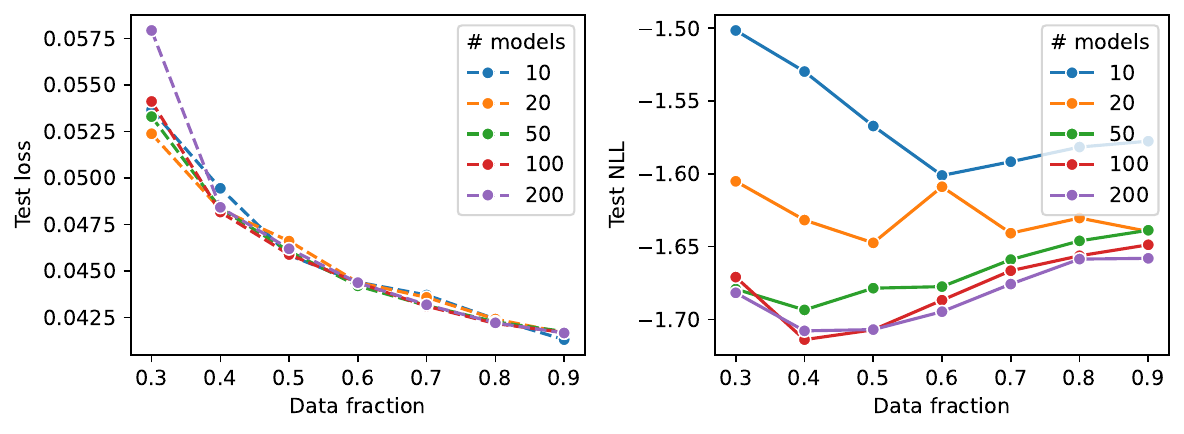}
    \caption{Hyperparameter exploration of model ensembles. Test loss and test negative log likelihood (NLL) for various data fractions and numbers of models.}
    \label{fig:ensemble-calibration}
\end{figure}

When training ensemble models, we need to choose 2 hyperparameters: the number of models in the ensemble and the fraction of the data that is seen by each model.
We carry out a hyperparameter sweep for MLP ensemble in which we include 10, 20, 50, 100, 200 models and the data fractions 0.3, 0.4, ..., 0.9.
In Figure~\ref{fig:ensemble-calibration} we plot the test loss and test negative log likelihood for this parameter sweep.
A few interesting observations emerge:
\begin{enumerate}
    \item Based on test NLL, it is beneficial to include more models, but the gains are negligible beyond 50 models. Therefore, we select 50 models for the ensembles.
    \item As the data fraction fed to each model increases,  test loss continues to reduce. However, variance between model prediction reduces (models become increasingly confident) which is detrimental to NLL predictions. 
    We select a data fraction of 0.33.
\end{enumerate}

\subsection{Simulated pool-based active learning}
All available data points are split into 2 pools: training pool and candidate pool. 
The models are calibrated using the data from the training pool and the next candidate is proposed from the candidate pool based on the \textit{expected improvement} acquisition function.
Since the data were acquired based on triplets, we follow the same principle in this method.
At the start, the training pool contains 3 triplets. The candidate pool is always evaluated based on triplets. 
Note that 3 out of 5 input variables are identical for all points in the triplet. 
The remaining 2 variables (catalyst mass loading, zero porosity thickness) are averaged over the three data points in the triplet.
Once a triplet is selected, all three points with their respective individual values of input variables are included in the training pool for the next round of model calibration.

Acceleration factor is defined as the ratio between the expected number of optimization steps with random candidate proposal and the corresponding expected number with the given strategy.
Note that for a candidate pool size of \(n\), the expectation for random proposal is \((n-1)/2\). 
We operate with an initial candidate pool size of \(n=27\). Therefore, \(AF=13/\mathbb{E}[N]\) where \(\mathbb{E}[N]\) is the mean number of steps needed to get to the maximum FE and for the given strategy.

\end{document}